\documentstyle[twocolumn,aps]{revtex}
\begin{document}
%\draft
%\preprint{Ames Laboratory-USDOE Preprint }
\title{
 Tight-Binding Parameterization for Photonic Band Gap Materials
}
\author{
E.~Lidorikis$^1$, M.~M.~Sigalas$^1$, E.~N.~Economou$^2$ and 
C.~M.~Soukoulis$^{1,2}$
}
\address{
$^1$Ames Laboratory-USDOE and Department of Physics and Astronomy,   
Iowa State University, Ames, Iowa 50011 \\
}
\address{
$^2$Research Center of Crete-FORTH and Department of Physics,   
University of Crete,Heraklio, Crete 71110, Greece \\
}
\author{\parbox[t]{5.5in}{\small
The ideas of the linear combination of atomic orbitals (LCAO) method,
well known from the study of electrons, is extended to the classical
wave case. The Mie resonances of the isolated scatterer in the classical
wave case, are analogous to the localized eigenstates in the electronic
case. The matrix elements of the two-dimensional tight-binding (TB) 
Hamiltonian are obtained by fitting to {\em ab initio} results. The
transferability of the TB model is tested by reproducing accurately the
band structure of different 2D lattices, with and without defects, thus
proving that the obtained TB parameters can be used to study other
properties of the photonic band gap materials. 
\\ \\
PACS numbers: 42.70Qs, 41.20Jb, 71.15-m}}
\maketitle
\normalsize 
%\tightenlines
%\newpage

In recent years experimental and theoretical studies of artificially
manufactured periodic dielectric media called photonic band gap (PBG)
materials or photonic crystals, have attracted considerable attention
\cite{bib1,bib2}. PBG materials can have a
profound impact in many areas in pure and applied physics.
PBG materials are often considered as analogous to electronic
semiconductors. The existence or not of spectral gaps in periodic
PBG materials  or localized states in disordered systems, in analogy
with what happens to the electronic materials, is of fundamental
importance. While it is by now firmly established that certain periodic
arrangement of dielectric structures has a full PBG in 2D and 3D
\cite{bib1,bib2}, it is not clear what mechanism is responsible for the
formation of gaps. The relative importance of the roles of the two
different mechanisms, single scatterer resonances and macroscopic
Bragg-like resonances, in the formation of gaps and localized states is
still being debated. Preliminary results \cite{bib6,bib7} have shown
that there is a direct correspondence between the gaps calculated by
plane wave expansion and the Mie resonances \cite{bib8} of an isolated
sphere. It is surprising that the positions of the Mie resonances
approximately coincide with the center of the bands. It is tempting to
suggest that the Mie resonances of an isolated scatterer to play the
role of the energy levels of an isolated atom in a crystal. Also, it
would be possible to formulate the problem in a simpler way, similar to
the tight-binding (TB) formulation of the electronic problem.   

It is well known that the TB method has  proven to be  very useful in
studying the electronic properties of solids \cite{slater,papaco,harr}.
In an empirical TB
approach, matrix elements of the Hamiltonian between orbitals centered
on different sites are treated as parameters which are adjusted to
obtain the band structure and the band gaps, which have been determined
by other more accurate methods. The parameters obtained in this way are
then used to study other properties of the systems, such as surface
states, impurities and properties of disordered systems. The success of
the TB formulation has been tested in the studies of Si, C and
hydrogenated amorphous systems \cite{papaco,tang}.

In this paper, we show that it is possible to extend the ideas of the
LCAO method to the classical wave case. The Mie resonances of the
isolated scatterer in the classical case play the role of the localized
eigenstates in the electronic case. However, there exist two important
differences. First, Mie resonances' states are not
localized, in fact they decay too slowly, as $1/r$ as $r\rightarrow
\infty$ and this may lead to divergences in some matrix elements. 
However, in a lattice environment they become localized, with a
localization length analogous to the interparticle dimension.
Second,
in the classical wave case, as opposed to the electronic case, the host
medium supports propagating solutions for every frequency. For large
wavelengths, this is the dominant propagation mode since no resonances
have been excited yet, while for wavelengths comparable to the particle
dimension, transmission is achieved through coupling with the fully
excited localized resonances. In that sense, we may assume that a plane
wave is coupled with the 
lowest frequency band, while all higher bands are described solely by
the higher excited resonances. This assumption is mostly
justified in the case of wide gaps and narrow bands, which is the one
that we will study. 
We find that under appropriate rescalings, all  matrix elements scale
well  with distance, and we
obtain their functional dependence. This TB parameterization 
of PBG materials will be very useful in studies of impurities and the
effects of disorder in these materials.

We will contend ourselves in the  scalar wave equation
case of a  2-D  periodic array of N infinitely long  dielectric
cylinders,  with 
periodic boundary conditions and the
incident plane wave E-polarized. We assume the normalized electric field
for  each
band to be given by
\begin{eqnarray}
E_n(\vec{r},\vec{k})=\frac{c_n^1(
k)}{\sqrt{N}} e^{i\vec{k}\vec{r}}
+ \frac{c^2_n(k)}{\sqrt{N}}\sum_{\vec{R}}\Psi_n(\vec{r}-\vec{R})e^{i\vec{k}\vec{R}} 
\end{eqnarray}
where $n=0,1,2,...$ is the resonance's  (or band) index, 
$\Psi_n(\vec{r}-\vec{R})$ stands for the wavefunction of the
$n$'th resonance localized at $\vec{R}$,  and has an
angular symmetry $\Psi \sim \cos (n\theta)$. \ $c_n^1=0$, $c^2_n=1$ for $n\neq 0$ and  are functions of the
frequency ($k\equiv \vert\vec{k}\vert$) with $\vert c_n^1 \vert ^2 +\vert c_n^2 \vert ^2=1$, \ \ $\vec{r}$, $\vec{R}$, $\vec{k}$ are 2-D vectors, and we have
assumed a unit area unit cell. We
will assume a  frequency independent normalized resonance function
$\Psi$, with
$\int\Psi_m^{\ast}(\vec{r})\Psi_n(\vec{r}-\vec{R})d\vec{r}=\delta_{mn}
\delta(\vec{R})$, as well as
$\int\Psi_m^{\ast}(\vec{r}-\vec{R})e^{i\vec{k}\vec{r}}d\vec{r}=0$ so
that $\int E_m^{\ast}E_nd\vec{r}=\delta_{mn}$ 
in order to simplify the problem and make a  better
correspondence with the electronic case. This will turn to be a good 
approximation for our case of interest. For the lowest frequency band,
we should expect $c_0^1\rightarrow 1,\ c_0^2\rightarrow 0$ for $\vert
\vec{k} \vert \rightarrow
0$ and  $c_0^1\rightarrow 0,\ c_0^2\rightarrow 1$ for $\vert \vec{k}\vert
\rightarrow \vert \vec{G}\vert/2$.

The ``Hamiltonian'' for the scalar wave equation is 
$H=-\vec{\nabla}^2/\epsilon(\vec{r})$ and the eigenfrequencies
 $\omega^2/c^2$ of the
system can be found by diagonalizing  the Hamiltonian matrix  
  $H_{mn}=\int E_m^{\ast} H E_n d \vec{r}$.
Then :
\begin{eqnarray}
H_{00}=\vert c^1_0\vert^2 \frac{\vert \vec{k}\vert^2}{\langle \epsilon
\rangle}+\vert c^2_0\vert^2\left(\beta_{00}
+\sum_{\vec{R}}\gamma_{00}(\vec{R})e^{i\vec{k}\vec{R}}\right)
\end{eqnarray}
where we have neglected
crossing terms between the plane wave and the resonances (this is not
completely justifiable, but it can be argued that they are not important for
frequencies close to the gap, and so  for impurity and disorder
studies). 
$\beta_{00}=\int
\Psi_0^{\ast}(\vec{r})H
\Psi_0(\vec{r})d\vec{r}$\ and\  $\gamma_{00}(\vec{R})=\int
\Psi_0^{\ast}(\vec{r})H
\Psi_0(\vec{r}-\vec{R})d\vec{r}$, and  $\langle \epsilon
\rangle$ is the averaged dielectric constant. We argue that  the
functional form of $\vert c_0^2(k)\vert^2$ is  similar to
the  form of the
scattering cross section of a single cylinder for the $n=0$ (or
$s$-wave) case, so that $ \vert c^1_0(k)\vert^2 \simeq   e^{-\lambda(f)
\omega_r^\mu}$. Here  $\omega_r=\vert \vec{k}\vert c/(\omega_0 \sqrt{\langle \epsilon
\rangle})$,  $\omega_0$ is the single cylinder Mie resonance frequency,  
$\lambda(f)$ is a function of the filling ratio $f$ of the form
$\lambda(f)=b_1/f^{b_2}$. The power $\mu$ has to be larger than $2$
in order to preserve the correct slope at $\vert \vec{k}\vert
\rightarrow 0$. For simplicity, we choose $\mu=4$.

The second band ($n=1$ or $p$-like) has a $\Psi \sim \cos \theta$
symmetry, and will consist of two linearly independent polarizations,
$p_x$ and $p_y$. There will be crossing terms between these two
polarizations, but not between $s$-like and $p$-like resonances. Using
the standard notation\cite{slater,papaco,harr} for the matrix elements, and
taking into account
only first and second nearest neighbors, then the nonzero Hamiltonian matrix
elements for the case of a square lattice with lattice constant $a$ is:
\begin{eqnarray}
H_{ss} & = & \vert c^1_0 \vert^2 \vert k \vert^2/\langle \epsilon \rangle
+\vert c^2_0\vert^2 [ \varepsilon_s
+2V_{ss\sigma}^{(1)} \times{} \nonumber\\
 {} & & \times (\cos \phi_x
+\cos \phi_y)
+4V_{ss\sigma}^{(2)}\cos\phi_x \cos\phi_y]\\
H_{p_xp_x} & = &\varepsilon_{p_x} +2V_{pp\pi}^{(1)}\cos
\phi_y+2V_{pp\sigma}^{(1)}\cos\phi_x +{} \nonumber\\ {} & & 
+2\left( V_{pp\sigma}^{(2)} +V_{pp\pi}^{(2)}\right)
\cos\phi_x\cos\phi_y\\ 
H_{p_xp_y} & = & 2\left( V_{pp\pi}^{(2)} -
V_{pp\sigma}^{(2)}\right)
\sin\phi_x\sin\phi_y
\end{eqnarray}
where  $\phi_x=k_x a$, $\phi_y=k_y a$ and  $\vert k \vert =\sqrt{k_x^2
+k_y^2}$. $V^{(1)}$, $V^{(2)}$  refer to the  $\gamma$
matrix  elements for first and second nearest neighbors respectively,
 $\varepsilon$ refers to
the  $\beta$, on-site or diagonal,
matrix element,  $H_{p_yp_y}$ is similar to $H_{p_xp_x}$ with
$x\longleftrightarrow y$ and  $H_{p_yp_x}=H_{p_xp_y}^*$. In this work we consider
only these two bands. 

We have fitted the $V$ and $\varepsilon$ matrix elements for the E
polarization case with cylinders of dielectric constant $\epsilon=100$
in vacuum, to the band
structure of five different rectangular lattices with large/small axis
ratios: $1,\ 1.05,\ 1.1,\
1.15,\ 1.2$ as well as to a hexagonal lattice, for six different
filling ratios: $f=.1,\ .2,\ .3,\ .4,\ .5,\ .6$. 
We used a large value for the dielectric constant to ensure that we
obtain large gaps and narrow bands.  The quality of these
fits can be seen in Fig. 1 where we plot the bands as found numerically
by the plane wave expansion (PWE) method along with the TB fit, for a square
and a hexagonal lattice for two filling ratios. The perfect quality of the
fits is an indication of the potential usefulness of the TB
method. 

We next plot some of the fitted matrix elements. The square root of the 
diagonal $\varepsilon_{p_x}$ and $\varepsilon_{p_y}$ matrix
elements are plotted (Fig.~2a) as a function of the filling ratio f, while
the off 
diagonal $V_{pp\pi}$ matrix elements are plotted (Fig.~2c) as a function
of  the dimensionless
separation  distance $d_{ij}=r_{ij}/\alpha$, where $r_{ij}$ is the
separation distance between cylinders $i$ and $j$ and $\alpha$ is the
cylinders' radius. Notice how bad the scaling is, especially for the
$V_{pp\pi}$ element. Apparently, a satisfactory 
TB description  can not be achieved with this two-center
approach, but rather, with the  inclusion of the lattice environment
\cite{tang}  as well.

The proposed simple rescaling function $(D_n^{on})_i$ for the diagonal
matrix element  that describes the lattice
environment of cylinder $i$, and
takes into account the filling ratio and the different symmetries is
of the form:
\begin{eqnarray}
\frac{1}{(D_n^{on})_i}=
\sum_{j\ne i} \frac{\tau \cos^2(n\theta_{ij})}{d_{ij}^{\nu_n}}
\end{eqnarray} 
where  $\theta_{ij}$ is the angle between the symmetry
axis of the $p$ resonance on cylinder $i$ and the $\hat{r}_{ij}$ direction,
  $n$=0,1,.. for the $s$,$p$,.. resonances, and the sum
runs over the nearest neighbors of cylinder $i$.  
The power on
the angular function was chosen so that the $p_x$ and $p_y$ resonances in the
hexagonal lattice be the same. The only choices were 2 and 4, and it was
found that 2 gives better results. Eq.~6 is similar to what was used in
Ref.~\cite{tang} for the atomic orbitals, except for two differences:
(a) here we take into account the resonance's angular symmetry, and (b)
the exponentially decaying part is missing, reflecting the non-localized
character of the EM resonances.  
Finally, $\tau=\left[ \pi/(a^2 f)\right]^2$ is a  renormalization  that takes
into account that different structures, with the same lattice constant
$a$ and cylinder radius $\alpha$, have different
filling ratios.  $\tau$ is normalized so that $\tau=1$ for the
rectangular structures and  $\tau=3/4$ for the hexagonal.
We will use this parameter only for the diagonal matrix elements.

For the periodic case, the function ($D_n^{on})_i$ is the same for every
$i$, and will characterize the corresponding resonance.
We find the diagonal
matrix element to scale as
 $\sqrt{\epsilon_n}=a^n_0+a_1^n (D_n^{on})^{-a_2^n}+a_3^n
(D_n^{on})^{-a_4^n}$ \ 
where $a^n_0=\omega_0 \alpha/c$ is the corresponding  dimensionless Mie
 resonance frequency. In Fig. 2b  we plot 
$\sqrt{\varepsilon_p}$ 
 {\em vs}  the environment function
$(1/D^{on}_p)$. We can see that $\sqrt{\varepsilon_p}$ now 
scales very well, having a larger value for increasing lattice density.
The same dependence is found for all bands.

 In order to rescale the off diagonal $V$ matrix element between 
two neighboring resonances $i$ and $j$, we need contributions from
the neighbors that are close to the line joining $i$ and
$j$. Contributions  have to be projected
on the $\hat{r}_{ij}$ direction for the $s$ resonance, while for
the $p$ resonance we have to project on its symmetry axis. Only first
nearest neighbors will contribute.   At the end, we have to normalize
with  the sum of all projection weights.

A simple formula that describes the environment of the $n$
resonance on cylinder $i$, along the $\hat{r}_{ij}$ direction is:
\begin{eqnarray}
\frac{1}{(D_n^{off})_{ij}}= 
\frac{\sum_l (\cos^2 \theta_{ilj}^n/d_{il}
^{\nu_n})}{\sum_l \cos^2 \theta_{ilj}^n}
\end{eqnarray}
where $l$ runs over $i$'s nearest neighbors (including $j$). 
$\theta^n_{ilj}$ is the angle
between the $\hat{r}_{il}$ and
$\hat{r}_{ij}$ directions for the $s$ resonance ($n=0$), and for the $p$
resonance  ($n=1$),
it is the angle between 
the $i$'th resonance's symmetry
axis and the $\hat{r}_{il}$ direction .
 Both angles are taken to have a range from $-\pi/2$ to  
$\pi/2$ \cite{ftn1}. 
Finally, if we include screening in our considerations, then the actual
matrix element $V$ to be used in a particular problem, can be obtained by
the fully rescaled one $\mathcal{V}$, by 
$V^{ij}=\mathcal{V}$$^{ij}(1-S^{ij})/[
(D^{off})_{ij}^{-1}+ (D^{off})_{ji}^{-1}]$\  
where $S^{ij}$ is the same  screening function used in Ref.~\cite{tang}:
$S^{ij}=\tanh \left( b_1 \sum_{l\ne
i,j}e^{-b_2\left[(d_{il}+d_{jl})/d_{ij}\right]^{b_3}}\right)$, and is
different for different matrix elements.
The fully rescaled matrix elements are found to scale with separation
distance as 
$\mathcal{V}$$^{ij}=c_1d_{ij}^{-c_2}+c_3d_{ij}^{-c_4}$. 
In Fig. 2d we plot the rescaled $\mathcal{V}$$_{pp\pi}$ matrix
element. We see now 
that it is a smouth function of separation distance, except for the
second nearest matrix elements  which 
do not scale very well for large filling ratios (small distances).
Apparently a screening function that depends on the filling ratio as
well is needed, but for simplicity and to keep some accordance
with the electronic models used so far,  we will use this one.

We find that
all matrix elements now scale very well with separation distance, so a
satisfactory TB description of PBG  materials
is in order. The rescaling parameters used  and the parameters for the 
rescaled matrix element curves' generation  are presented
in  Tables 1 and 2. We have also found $\nu_n=1.65$ for all $n$, and
$b_1=.068$, $b_2=1.23$ for $\lambda(f)$.
\begin{center}
TABLE I. The parameters for the $\varepsilon$ elements.\\
\begin{tabular}{|c||c|c|c|c|c|}
\hline
 & $a_0$ & $a_1$ & $a_2$ & $a_3$ & $a_4$\\
\hline \hline
$\ \sqrt{\varepsilon_s}\ $ & 0.0804 & 0.0460 & 0.716 & -0.0121 & 5.000\\
\hline
$\ \sqrt{\varepsilon_p}\ $ & 0.2371 & 0.0890 & 1.640 & 0.0020 & 0.320 \\
\hline
\end{tabular}
\end{center}
\begin{center}
TABLE II. The parameters for the $V$ elements.\\
\begin{tabular}{|c||c|c|c|c|c|c|c|}
\hline
 &  $b_1$ & $b_2$ & $b_3$& $c_1$ & $c_2$ & $c_3$ & $c_4$ \\
\hline \hline
$V_{ss\sigma}$ & 0.075 & 0.0008 & 10.0 & -0.108 & 4.00 & -0.00096 & 1.30\\ 
\hline
$V_{pp\sigma}$ & 0.100 & 0.00008 & 13.5 & 0.425 & 6.14 & 0.0550 & 3.22 \\
\hline
$V_{pp\pi}$ & 0.700 & 0.0015 & 10.0 & -0.076 & 5.40 & -0.0044 & 2.32\\
\hline
\end{tabular}
\end{center}

Next, we test our results on a nonperiodic lattice. The original
periodic lattice is  square with filling ratio $f=0.2$ and lattice
constant $a$. We will solve
the  3$\times$3 supercell problem. Each supercell consists of 9
cylinders, and the whole lattice is periodic in terms of this supercell.
We keep all cylinders in the supercell in place, except the middle one,
which we move from it's periodic position $(0,0)$, $a/4$ 
 towards the left, then $a/4$ towards the bottom,
and then return to the periodic position $(0,0)$ along the diagonal (see
insert graph in 
Fig. 3). Solving the supercell TB problem involves a 9$\times$9 matrix
diagonalization for the $s$ band, and a 18$\times$18 matrix
diagonalization for the $p$ band. We plot the 3 highest eigenvalues
of the $s$ band, and the 3 highest and 3 lowest eigenvalues of the $p$
band. Luckily, for the 3 highest eigenfrequencies of the $s$ band, we
do not need to take into account the free plane wave, since for those
frequencies the resonances are fully excited, blocking the background
propagation mode.
We plot the eigenfrequencies for $\vec{k}a=(0,1/3)$, and directly
compare our results with the ones obtained
numerically by the PWE method, for the same exactly system.
It is worth noting that in the PWE method, 100 plane waves per rod are
needed for an acceptable numerical convergence. This is a factor of
$10^4$ more CPU time than the TB method.  

The agreement is excellent, so our TB parameterization works very well
for the impurity case too. The photonic TB model proves to be
an excellent tool for understanding and describing the band structure and
properties of light in PBG structures. We hope that the calculated TB
parameters  will be used in studies of 2D PBG materials and can
stimulate enough interest to extend these studies in 3D photonic crystals.

We would like to thank C.~Z.~Wang and K.~M.~Ho for useful discussions.
Ames Laboratory is operated for the U.S. Department of Energy by Iowa
State University under Contract No. W-7405-Eng-82. This work was supported by 
the Director for Energy Research office of
Basic Energy Sciences and Advanced Energy Projects,  by NATO Grant 
No. 940647, and by a $\Pi$ENE$\Delta$ grant.

\begin{figure} 
\caption{The first two frequency bands for a square lattice with filling
ratio $f=.1$ (a) and $f=.4$ (b), and for a hexagonal
lattice with $f=.1$ (c) and $f=.4$ (d).  Circles
correspond to numerical results from the PWE method,
while solid lines correspond to the TB fit. }
\end{figure}
\begin{figure} 
\caption{Fitted TB parameters for the second ($p$-like) frequency
band. (a) $\sqrt{\varepsilon_p}$ {\em vs} $f$. 
(b)  $\sqrt{\varepsilon_p}$ {\em vs} the rescaled environment function
 $1/D^{on}_p$. 
Circles and squares correspond to $\sqrt{\varepsilon_{p_x}}$ and
$\sqrt{\varepsilon_{p_y}}$ respectively of a rectangular lattice with
the big axis along $\hat{x}$, and triangles to a hexagonal lattice.
(c) $V_{pp\pi}$ {\em vs} the separation $d$.
(d) The rescaled $\mathcal{V}$$_{pp\pi}$ {\em vs} 
$d$. Circles, squares and diamonds  correspond to a rectangular
lattice's $V_{pp\pi}$
elements along the small axis, the large axis and the diagonal, 
and triangles  to a hexagonal lattice.    
All matrix elements are expressed in
the dimensionless units of $(\omega \alpha /c)^2$.} 
\end{figure}
\begin{figure} 
\caption{The three edge eigenfrequencies for each band for a $3 \times
3$ impurity system. All cylinders are identical, with the midlle one
moving from the equilibrium position, as pointed in the insert
graph. The wave vector is always constant $\vec{k}a=(0,1/3)$. The first band
gap extends approximately from $\omega a /c \simeq .53$ to $\omega a /c
\simeq .87$,
while the second starts at $\omega a /c \simeq 1.03$}
\end{figure}

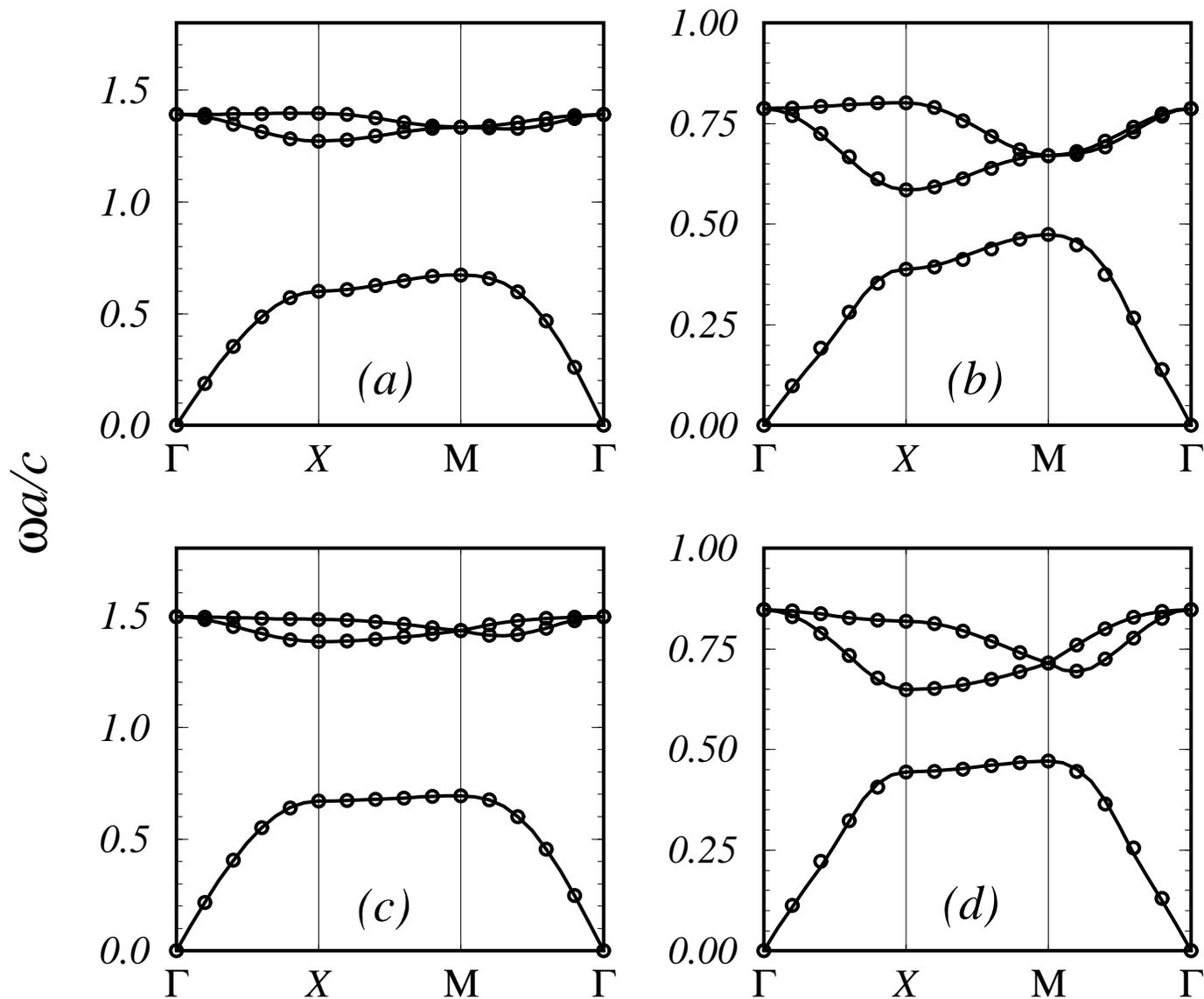
\end{document}